# Observation of topologically enabled complete polarization conversion

Fujia Chen[1, 2], Zhen Gao[3, *], Li Zhang[1, 2], Qiaolu Chen[1, 2], Qinghui Yan[1, 2], Rui Xi[1, 2], Liqiao Jing[2], Erping Li[2], Wenyan Yin[1, *], Hongsheng Chen[1, 2, *], Yihao Yang[1, 2, *]

[1]Interdisciplinary Center for Quantum Information, State Key Lab. of Modern Optical Instrumentation, College of Information Science and Electronic Engineering, Zhejiang University, Hangzhou 310027, China.

[2]ZJU-Hangzhou Global Science and Technology Innovation Center, Key Lab. of Advanced Micro/Nano Electronic Devices & Smart Systems of Zhejiang, ZJU-UIUC Institute, Zhejiang University, Hangzhou 310027, China.

[3]Department of Electrical and Electronic Engineering, Southern University of Science and Technology, Shenzhen 518055, China.

*E-mail: yangyihao@zju.edu.cn (Y.Y.); gaoz@sustech.edu.cn (Z.G.); hansomchen@zju.edu.cn (H. C); wyyin@zju.edu.cn (W.Y.)

**Exploiting topological ideas has been a major theme in modern photonics, which provides unprecedented opportunities to design photonic devices with robustness against defects and flaws[1,2]. While most previous works in topological photonics have focused on band theory, recent theoretical advances extend the topological concepts to the analysis of scattering matrices[3-6] and suggest a topological route to complete polarization conversion (CPC)[3-5], a unique photonic phenomenon without an electronic counterpart. Here, we report on the experimental observation of the topological effect in reflection matrices of a photonic crystal slab, enabling CPC between two linear polarizations over a wide range of frequencies. Using angle-resolved reflection measurements, we observe CPC occurring at vortex singularities of reflection coefficients in momentum space, verifying the topological nature of CPC. In addition, the topological effect also guarantees the spin-preserved reflection of a circularly polarized wave. Remarkably, we experimentally establish a connection between two seemingly unrelated topological phenomena—CPC and bound states in the continuum (BICs)[7]: BICs lie on the critical coupling curves that define the condition for CPC. Our work paves the way to exploring the topological properties in scattering matrices for controlling light polarization and spin and creating robust photonic devices.**

Topology is a mathematical concept concerned with geometric properties that are preserved under continuous deformation. In recent years, there has been a rapidly growing interest in exploiting topological properties of physical systems characterized by quantized topological invariants. In a topologically nontrivial system, the physical quantities associated with topological invariants are robust against small perturbations, as the quantized invariants cannot change continuously. In photonics, topological ideas are currently under active investigation to design and control the behavior of light. Previous works on topological photonics have largely focused on topological band theory and topological invariants (such as Chern number) defined by Bloch wavefunctions in photonic band structures[1,2,8-15]. As a physical consequence of the band topology, backscattering-immune edge states reside at interfaces between two topologically distinct media, enabling a plethora of applications ranging from robust optical delay lines[16] to on-chip communications[17] and topological lasers[18-20].

Recently, tremendous efforts have been devoted to achieving topological photonic phenomena beyond the framework of topological band theory and with no parallel in condensed-matter physics. For instance, it has been revealed that the far-field polarization vector of guided resonance modes of photonic crystal (PhC) slabs can exhibit vortex configurations in momentum space, which carry quantized topological charges defined by the number of times of polarization vectors winding around the vortex singularity. Intriguingly, the guided resonance mode at the vortex singularity has an infinitely large quality factor, also known as a bound state in the continuum (BIC)[7,21-23]. Moreover, owing to their topological nature[24], BICs are robust against small perturbations and have found many applications, including high-quality-factor lasers[25], optical vortex generators[26-28], and ultrasensitive sensors[29].

More recently, topological concepts have been theoretically extended to the analysis of scattering matrices[3-6]. This shows the possibility of realizing the complete polarization conversion (CPC) between two linear polarizations and the handedness-preserved reflection of a circularly polarized wave over a wide range of frequencies at isolated incident wavevectors, which is enabled by the topological properties of scattering matrices[3]. This new mechanism not only deepens our understanding of scattering matrices but also shows a promising future toward multiple applications, including high-efficiency polarization conversion[3], spin-conserved mirrors[3], arbitrary polarization conversion[4], and all-optical ultrafast polarization modulation when involved with nonlinearity[5]. However, experimental demonstration of the topological effect in scattering matrices as well as the topologically enabled CPC has remained an outstanding task due to the lack of an ideal photonic system.

Here, we report the direct experimental observation of the topological effect in scattering matrices and the topological CPC between two linear polarizations over a wide range of frequencies, as the electromagnetic wave is reflected from a simple PhC slab (see Fig. 1**a**). Our designed PhC features a "clean" band structure of the guided resonances—only a single band in the frequency range of interest, enabling CPC to span over the entire light cone, in contrast to a small portion of the light cone in the previous proposals[3-5]. Therefore, our PhC slab serves as an ideal platform to study the topological effect in reflection matrices and the topological CPC. Using polarization-sensitive angle-resolved reflection measurements, we directly probe the vortex configuration and the topological charge of the complex reflection coefficients in momentum space. We experimentally confirm that CPC occurs at vortex singularities at isolated incident momenta, implying the topological nature of CPC. In addition, the topological effect also guarantees the spin-preserved reflection of a circularly polarized wave, in stark contrast to the spin-flipped reflection in usual mirrors (see Fig. 1**a**). Remarkably, we experimentally identify an intriguing connection between two distinct topological phenomena—the topologically protected CPC and the topologically enabled BICs—and demonstrate that BICs lie on the critical coupling curves that define the condition for CPC.

As illustrated in Fig. 1**a**, our structure is the well-known "Sievenpiper mushroom" PhC[30] that consists of an array of square copper patches with side length $b = 8$ mm and lattice constant $a = 10$ mm. These patches are separated from the copper ground plane by a dielectric substrate [F4B ($\varepsilon = 3.5 + 0.007i$)] with thickness $h = 4$ mm, and each copper patch connects to the ground plane through a metallic via with radius $r = 2$ mm. Only zeroth-order diffraction can occur, as the frequency range of interest is below the diffraction limit $f_c = c/a = 30$ GHz ($c$ is the speed of light in free space). Consequently, the reflection matrix of our designed reflective PhC slab can be defined as $R = [R_{ss}, R_{ps}; R_{sp}, R_{pp}]$, where $R_{\sigma\mu}$ denotes the reflection coefficients of a $\mu$-polarized incident wave reflected into a $\sigma$-polarized wave with $\sigma, \mu \in \{s, p\}$. Here, the polarization is defined with respect to the incident plane composed of the incident wave vector $\hat{k}$ and the normal vector of the PhC slab $\hat{z}$. For *p*-polarized waves, the electric field (the magnetic field) is parallel (perpendicular) to the incident plane; for *s*-polarized waves, the electric field (the magnetic field) is perpendicular (parallel) to the incident plane.

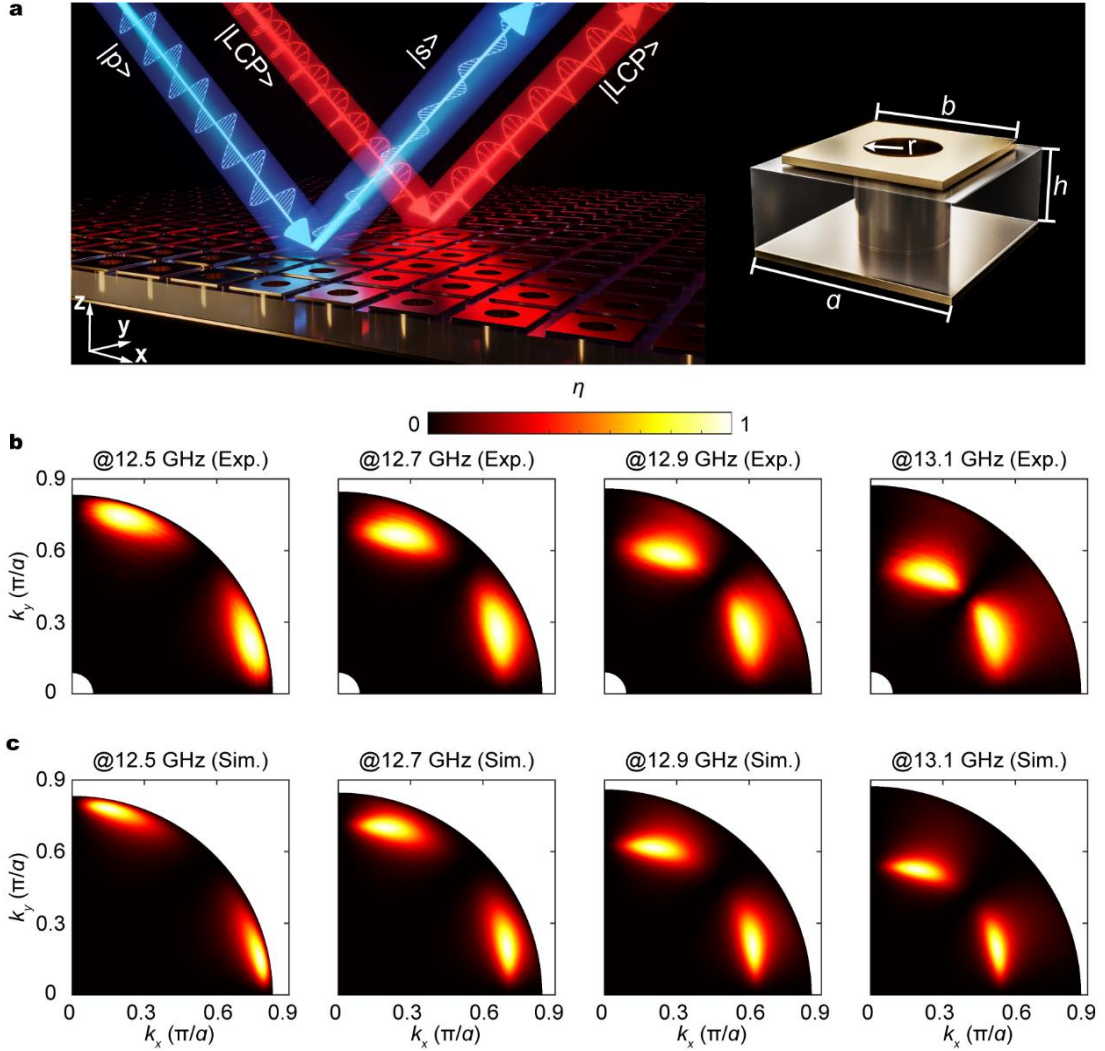

**Fig. 1 | Experimental observation of the topologically enabled CPC. a,** Schematic of the experiment setup. The designed PhC slab consists of an array of square copper patches connected to the ground plane. The PhC slab can provide CPC between two linear polarizations and total reflection of a circularly polarized wave without flipping its handedness. |s>, |p>, and |LCP> represent *s*-, *p*-, and left-handed circularly polarized waves, respectively. The right inset shows the details of the unit cell. Here, $a$ = 10 mm; $h$ = 4 mm; $b$ = 8 mm; $r$ = 2 mm; the permittivity of the substrate is $3.5 + 0.007i$. **b,c,** Measured (**b**) and simulated (**c**) polarization conversion efficiency $\eta$ in momentum space at 12.5 GHz, 12.7 GHz, 12.9 GHz, and 13.1 GHz, respectively. CPC with $\eta = 1$ (the brightest areas) occurs at two isolated wave vectors ($k_x$, $k_y$) in a quarter of the *k*-space. The color represents the polarization conversion efficiency $\eta$.

In experiments, we adopt a pair of highly collimated lens antennas as transmitting and receiving antennas, and the corresponding wavevector of incident waves parallel to the PhC slab is $\boldsymbol{k}_{//} = (k_x, k_y)$

(see more details in Supplementary Information). All elements in $R$ can be experimentally measured by tuning the polarization of each antenna (see experimental details in Methods). Slight material loss is inevitable in experiments but has a negligible impact on our main conclusions, which has been discussed in detail in previous theoretical work[3]. To better reflect the topological nature of the cross-polarization conversion, we define its efficiency as $\eta = |R_{sp}|^2 / (|R_{sp}|^2 + |R_{pp}|^2)$, approaching $|R_{sp}|^2$ in the lossless limit. In this generalized definition, the complete cross-polarization conversion efficiency $\eta = 1$ is equivalent to $|R_{pp}|^2 = 0$, which is valid for both lossy and lossless cases. The measured $\eta$ as a function of $\boldsymbol{k}_{//}$ at different frequencies is shown in Fig. 1**b**. We observe CPC at two isolated $k$-points related by the mirror symmetry in a quarter of $k$-space from 12.3 GHz to 13.4 GHz (see more measured results in Fig. S1 of Supplementary Information). We also perform simulations and can see two $k$-points where CPC occurs at each frequency (see Fig. 1**c**). Therefore, the effect of CPC is robust against frequency variation.

Next, we experimentally prove that CPC is a topological effect tied to the zeros in $R_{pp}$. We measure $R_{pp}$ as a function of $\boldsymbol{k}_{//}$ at different frequencies and plot the vector field of (Re ($R_{pp}$), Im ($R_{pp}$)) in $k$-space. As shown in Fig. 2**a**, one can see a saddle point in the upper triangular region with $k_y > k_x$ and a drain point in the lower triangular region with $k_y < k_x$. Moreover, these two topological singularities are related by the mirror operation with respect to the $y = x$ plane. The topological charges carried by the singularities can be defined as

$$q = \frac{1}{2\pi} \oint_C \nabla_{\vec{k}} \Phi(\vec{k}) d\vec{k}, q \in \mathbb{Z}, \tag{1}$$

which describes the times of the vector field of (Re ($R_{pp}$), Im ($R_{pp}$)) winding around the singularity. Here, $C$ is a closed path around the singularity, traversed in the anticlockwise direction, and $\Phi$ is the phase of $R_{pp}$. In Fig. 2**c**, we plot the measured phase distribution of $R_{pp}$ in $k$-space, namely, $\Phi(k_x, k_y)$. One can see that the phase decreases anticlockwise by $2\pi$ around the saddle point, indicating a topological charge $q = -1$. Conversely, the phase increases anticlockwise by $2\pi$ around the drain point, indicating a topological charge $q = +1$. The reflection dip at 12.9 GHz in the co-polarization reflection spectrum (blue curve in Fig. 2**d**) at the drain point implies that $|R_{pp}|^2$ is indeed zero, and CPC occurs. The simulated results in Figs. 2**b**, **e**, and **f** also show a similar topological effect. Therefore, the measured nonzero topological charge associated with $R_{pp} = 0$ verifies that CPC is topological in nature.

We also demonstrate that at the CPC $k$-points, the handedness of a circularly polarized incident wave is preserved after reflection from PhC. We measured the energy ratio $\eta_{LL} = |R_{LL}|^2 / (|R_{LL}|^2 + |R_{RL}|^2)$ at the drain point, with $|R_{LL}|$ ($|R_{RL}|$) being the co-(cross-) circular polarization conversion coefficient of the left-handed circularly polarized (LCP) incident wave (see more details in Supplementary Information). The high energy ratio at the singularity ($\eta_{LL} > 94\%$ in experiments and $\eta_{LL} = 100\%$ in simulations; see the yellow curves in Figs. 2**d** and **f**) indicates that the handedness of the incident wave is indeed preserved, which is an unusual phenomenon that is utterly distinct from the flipped handedness in ordinary mirrors. More importantly, such a chirality invariance is also topologically protected.

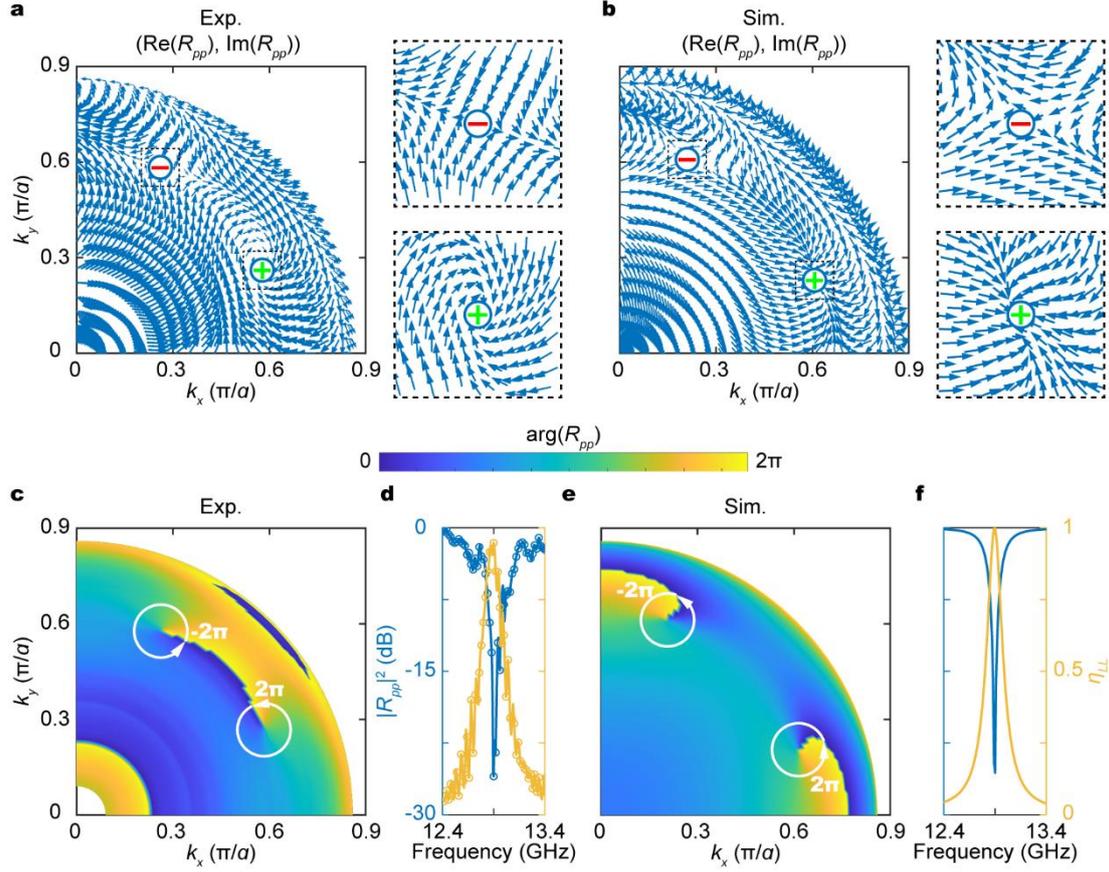

**Fig. 2 | Experimental demonstration of the topological nature of CPC. a,b,** Measured (**a**) and simulated (**b**) vector fields (Re($R_{pp}$), Im($R_{pp}$)) in *k*-space at 12.9 GHz with zoom-in plots near isolated *k*-points where $R_{pp} = 0$. **c-f,** Measured (**c,d**) and simulated (**e,f**) phase distributions of $R_{pp}$ in *k*-space at 12.9 GHz, co-polarization reflection spectra $|R_{pp}|^2$ at the phase singularity (blue curves), and energy ratio $\eta_{LL}$ at the phase singularity (yellow curves). Both the vector field and phase distribution of $R_{pp}$ exhibit nonzero topological charge, implying the topologically nontrivial nature of CPC and the spin-preserved total reflection.

To explain the underlying physical mechanism of the above intriguing phenomena, we measure the $|R_{pp}|^2$ and $|R_{ss}|^2$ spectra along the high-symmetry lines M-Γ-X (see experimental details in Methods). Due to the slight absorption loss in our structure, the reflection dips (the dark regions in Fig. 3**a**) indicate the guided resonances of the PhC slab (see Fig. 3**b**). Interestingly, the *p*-polarized incident waves cannot excite the guided resonance along ΓX (see Fig. 3**a**). In other words, the coupling constant of the guided resonance to *p*-polarized waves $|d_p| = 0$ for the wavevector along ΓX. Likewise, the coupling constant of the guided resonance to *s*-polarized waves $|d_s| = 0$ for the wavevector along ΓM (see Fig. 3**b**). Therefore, $|d_s| = |d_p| = 0$ at the Γ point, which indicates the existence of a BIC. BIC can also be manifested from the vanishing reflection dip around the Γ point in Figs. 3**a,b** and the numerically calculated infinitely large quality factor in Fig. 3**c**.

The zero coupling and the existence of a BIC can be understood intuitively from the perspective of mode symmetry. The colormaps, as insets in Fig. 3**c**, listed from right to left, are the $E_z$-field distributions of eigenmodes at wavevectors along ΓX, at the Γ point, and along ΓM, respectively. One can see that the eigenmode along ΓX is odd for the mirror plane $y = 0$. Therefore, the corresponding mode cannot couple to *p*-polarized waves, i.e., $|d_p| = 0$. Likewise, the eigenmode along ΓM is even for the mirror plane $y = x$, and thus, the corresponding mode cannot couple to *s*-polarized waves, i.e., $|d_s| = 0$. Finally, BIC at the Γ point has both symmetries and therefore cannot couple to both *s*- and *p*-polarized waves, i.e., $|d_s| = |d_p| = 0$. Such a BIC is also known as a symmetry-protected BIC[24]. We also numerically calculate the coupling constants to *s*- or *p*-polarized waves, indicating that $|d_s|$ and $|d_p|$ are zero along ΓM and ΓX, respectively (see Figs. 3**d,e**).

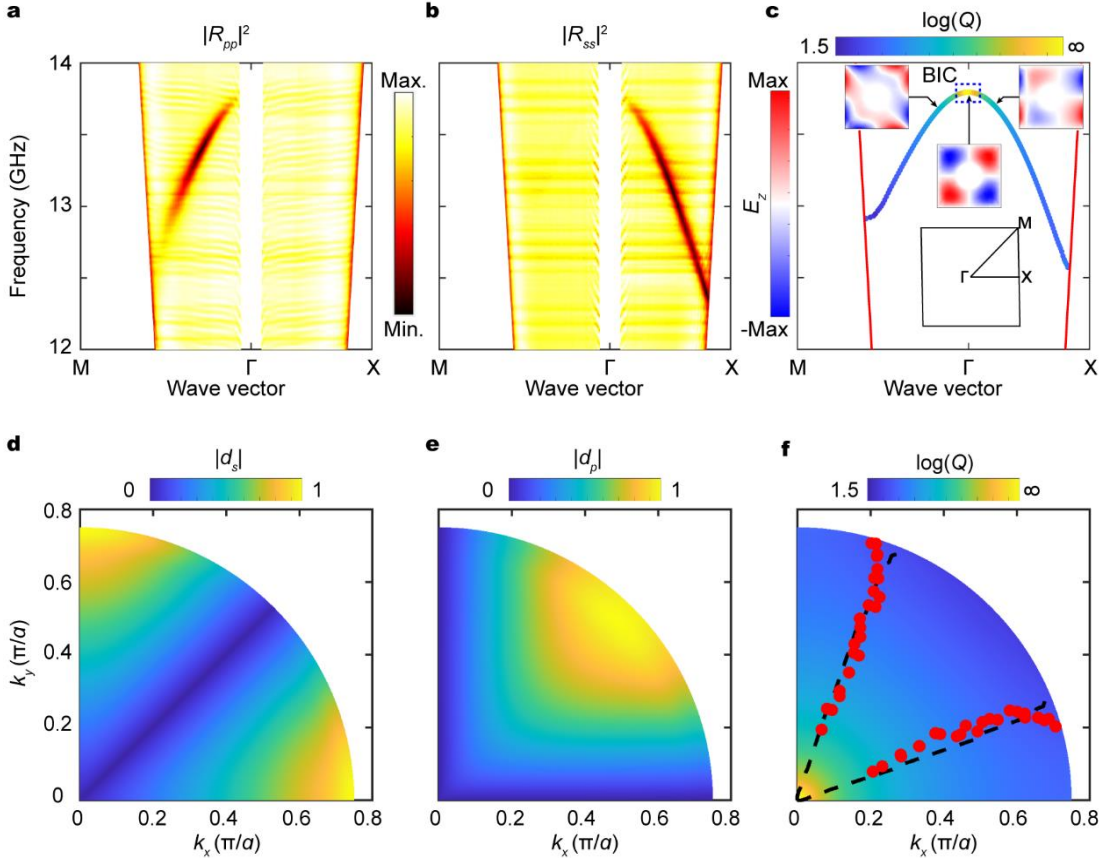

**Fig. 3 | Topologically protected CPC occurring at the critical coupling curves. a,b** Measured $|R_{pp}|^2$ (**a**) and $|R_{ss}|^2$ (**b**) along high-symmetry lines M-Γ-X. **c,** Simulated band structure of the guided resonance modes. The color represents the quality factors of the guided resonance modes on a logarithmic scale, approaching infinity at the Γ point. Insets: the $E_z$-field distributions of the eigenmodes (upper) and the first Brillouin zone (lower). The dashed box in **b** denotes the symmetry-protected BIC. **d,** Simulated $|d_s|$. $|d_s|$ is zero along ΓM. **e,** Simulated $|d_p|$. $|d_p|$ is zero along ΓX. **f,** Measured critical coupling curves. The red dots are extracted from the measured reflection spectrum $|R_{pp}|^2$ from 12.6 GHz to 13.7 GHz, indicating the CPC positions. The black dashed lines represent the numerically calculated critical coupling curves on which $|d_s| = |d_p|$.

According to previous theoretical work[3] (also see Supplementary Information), CPC occurs under critical coupling conditions, where $|d_p| = |d_s|$. Because $|d_p| = 0$ along ΓX and $|d_s| = 0$ along ΓM, there must be at least one critical coupling curve with $|d_s| = |d_p|$ residing between ΓX and ΓM[3]. Therefore, we can achieve CPC along this critical coupling curve unless $|d_s| = |d_p| = 0$. The numerically calculated critical coupling curves are plotted as the black dashed lines in Fig. 3**f**. We also experimentally extract the positions in the *k*-space where CPC occurs from the measured reflection coefficients of $|R_{pp}|^2$, which distribute on the critical coupling curves. Thus, we experimentally prove that the critical coupling defines the condition for CPC and that the symmetry-protected BIC lies at the critical coupling curves.

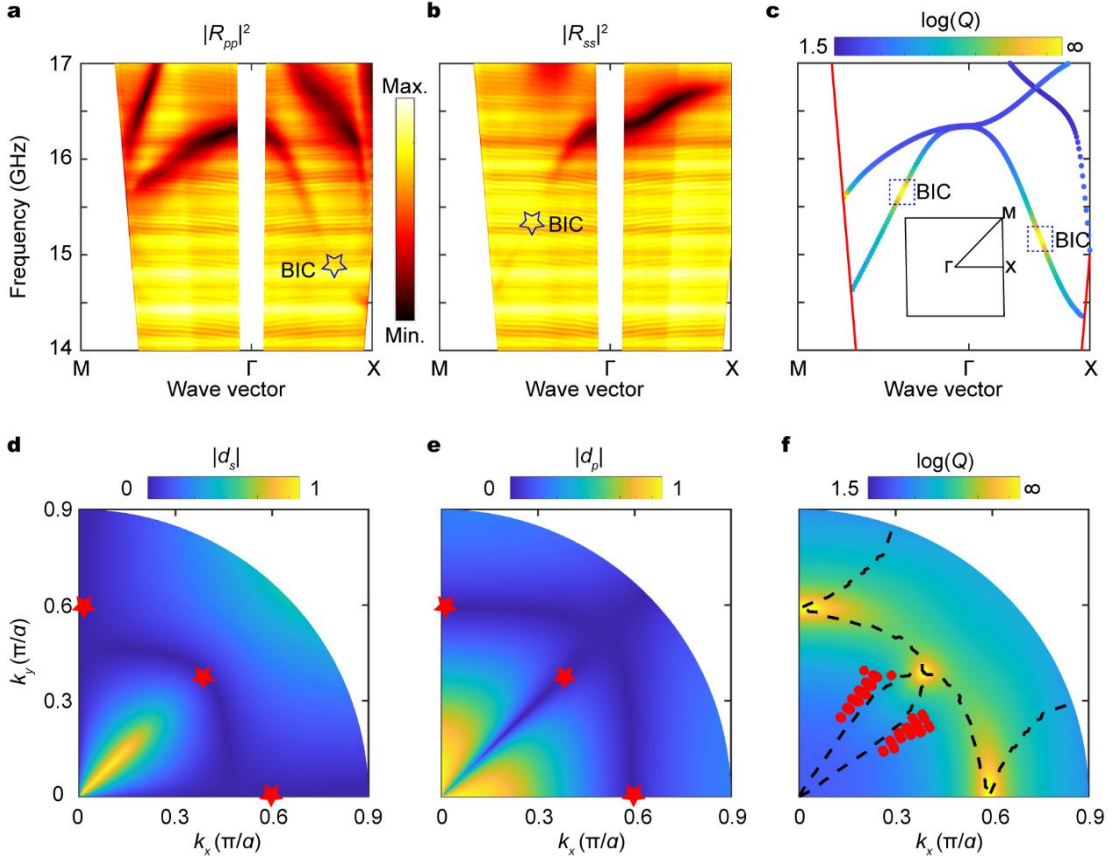

**Fig. 4 | Nonsymmetry-protected BICs and the critical coupling curves. a,b** Measured $|R_{pp}|^2$ (**a**) and $|R_{ss}|^2$ (**b**) along the high-symmetry lines M-Γ-X. The blue pentagrams denote nonsymmetry-protected BICs. **c,** Numerically calculated band structures of the guided resonance modes. The color represents the quality factors of the guided resonances on a logarithmic scale. Inset: the first Brillouin zone. **d,** Simulated $|d_s|$. $|d_s|$ is zero along ΓM except for the Γ point. **e,** Simulated $|d_p|$. $|d_p|$ is zero along ΓM except for the Γ point. The nonsymmetry-protected BICs are located at the red pentagrams. **f,** Measured critical coupling curves. The red dots are extracted from the measured reflection spectrum $|R_{pp}|^2$ from 15.7 GHz to 16.2 GHz. The background color represents the quality factors of the guided resonances on a logarithmic scale. The black dashed curves represent the numerically calculated critical coupling curves on which $|d_s| = |d_p|$. Here, the substrate thickness is $h = 5.3$ mm, and the remaining parameters are unchanged.

In addition to the symmetry-protected BICs, the nonsymmetry-protected BICs away from the Γ point[24] also lie on the critical coupling curves. To highlight this argument, we design and fabricate another sample with thickness $h = 5.3$ mm and keep the remaining geometrical parameters unchanged. Using similar measurements, we experimentally obtain $|R_{pp}|^2$ and $|R_{ss}|^2$ along M-Γ-X that exhibit vanishing reflection dips at two distinct positions away from the Γ point (see the blue pentagrams in Figs. 4**a**,**b**), matching well with the corresponding guided resonant modes exhibiting diverging quality factors (see the blue dashed boxes in Fig. 4**c**). Thus, these two points with vanishing reflection dips and diverging quality factors correspond to nonsymmetry-protected BICs.

From Fig. 4**a**, we can conclude that $|d_s| = 0$ along ΓX and $|d_p| = 0$ along ΓM for the lower band. This is consistent with the numerically calculated coupling constants shown in Figs. 4**d**,**e**. Note that due to the twofold band degeneracy, $|d_p| \neq 0$ and $|d_s| \neq 0$ at the Γ point. We also experimentally extract some *k*-points where CPC occurs (see the red circles in Fig. 4**f**). For comparison, the simulated critical coupling curves are plotted as dashed curves in Fig. 4**f**, with the background color representing the numerically calculated quality factors of the guided resonances. The red circles are distributed near the critical coupling curves, and the nonsymmetry-protected BICs reside at the critical coupling curves.

We have thus experimentally identified the topological effect in scattering matrices of PhC slabs, which gives rise to the topologically protected CPC and spin-conserved total reflection over a wide range of frequencies. In addition, we experimentally establish a deep connection between topologically protected CPC and BICs. Our work sheds light on exploring topological concepts beyond the conventional framework of band topology. We anticipate a richer set of intriguing topological phenomena when considering a higher dimensional space or the other aspects of scattering matrices. In terms of applications, our work shows promise for use in robust functional polarizers and spin-conserved mirrors. Our work also provides a powerful experimental technique to measure the critical coupling curves, which will significantly facilitate the experimental study and understanding of BICs. As the scattering matrices characterize the general responses of any linear wave system, it would be interesting to investigate physical phenomena and device applications based on the topological effects in scattering matrices of other wave systems, such as elastics, acoustics, and polaritons.

**References**

1      Lu, L., Joannopoulos, J. D. & Soljačić, M. Topological photonics. *Nat. Photon.* **8**, 821-829 (2014).

2      Ozawa, T. *et al.* Topological photonics. *Rev. Mod. Phys.* **91**, 015006 (2019).

3      Guo, Y., Xiao, M. & Fan, S. Topologically protected complete polarization conversion. *Phys. Rev. Lett.* **119**, 167401 (2017).

4      Guo, Y., Xiao, M., Zhou, Y. & Fan, S. Arbitrary polarization conversion with a photonic crystal slab. *Adv. Optical Mater.* **7**, 1801453 (2019).

5      Wu, Y., Kang, L., Bao, H. & Werner, D. H. Exploiting topological properties of Mie-resonance-based hybrid metasurfaces for ultrafast switching of light polarization. *ACS Photonics* **7**, 2362-2373 (2020).

6      Sakotic, Z., Krasnok, A., Alu, A. & Jankovic, N. Topological scattering singularities and embedded eigenstates for polarization control and sensing applications. *Photonics Res.* **9**, 1310-1323 (2021).

7      Hsu, C. W., Zhen, B., Stone, A. D., Joannopoulos, J. D. & Soljačić, M. Bound states in the continuum. *Nat. Rev, Mater.* **1**, 16048 (2016).

8      Haldane, F. D. & Raghu, S. Possible realization of directional optical waveguides in photonic crystals with broken time-reversal symmetry. *Phys. Rev. Lett.* **100**, 013904 (2008).

9. Wang, Z., Chong, Y., Joannopoulos, J. D. & Soljacic, M. Observation of unidirectional backscattering-immune topological electromagnetic states. *Nature* **461**, 772-775 (2009).

10. Fang, K., Yu, Z. & Fan, S. Realizing effective magnetic field for photons by controlling the phase of dynamic modulation. *Nat. Photon.* **6**, 782-787 (2012).

11. Khanikaev, A. B. *et al.* Photonic topological insulators. *Nat. Mater.* **12**, 233-239 (2013).

12. Rechtsman, M. C. *et al.* Photonic Floquet topological insulators. *Nature* **496**, 196-200 (2013).

13. Lu, L. *et al.* Experimental observation of Weyl points. *Science* **349**, 622-624 (2015).

14. Wu, L. H. & Hu, X. Scheme for achieving a topological photonic crystal by using dielectric material. *Phys. Rev. Lett.* **114**, 223901 (2015).

15. Yang, Y. *et al.* Realization of a three-dimensional photonic topological insulator. *Nature* **565**, 622-626 (2019).

16. Hafezi, M., Demler, E. A., Lukin, M. D. & Taylor, J. M. Robust optical delay lines with topological protection. *Nat. Phys.* **7**, 907-912 (2011).

17. Yang, Y. *et al.* Terahertz topological photonics for on-chip communication. *Nat. Photon.* **14**, 446-451 (2020).

18. Bahari, B. *et al.* Nonreciprocal lasing in topological cavities of arbitrary geometries. *Science* **358**, 636-639 (2017).

19. Bandres, M. A. *et al.* Topological insulator laser: Experiments. *Science* **359**, 1231 (2018).

20. Zeng, Y. *et al.* Electrically pumped topological laser with valley edge modes. *Nature* **578**, 246-250 (2020).

21. Hsu, C. W. *et al.* Observation of trapped light within the radiation continuum. *Nature* **499**, 188-191 (2013).

22. Jin, J. C. *et al.* Topologically enabled ultrahigh-Q guided resonances robust to out-of-plane scattering. *Nature* **574**, 501-504 (2019).

23. Yin, X. F., Jin, J. C., Soljacic, M., Peng, C. & Zhen, B. Observation of topologically enabled unidirectional guided resonances. *Nature* **580**, 467-471 (2020).

24. Zhen, B., Hsu, C. W., Lu, L., Stone, A. D. & Soljacic, M. Topological nature of optical bound states in the continuum. *Phys. Rev. Lett.* **113**, 257401 (2014).

25. Kodigala, A. *et al.* Lasing action from photonic bound states in continuum. *Nature* **541**, 196-199 (2017).

26. Doeleman, H. M., Monticone, F., den Hollander, W., Alù, A. & Koenderink, A. F. Experimental observation of a polarization vortex at an optical bound state in the continuum. *Nat. Photon.* **12**, 397-401 (2018).

27. Huang, C. *et al.* Ultrafast control of vortex microlasers. *Science* **367**, 1018-1021 (2020).

28. Wang, B. *et al.* Generating optical vortex beams by momentum-space polarization vortices centred at bound states in the continuum. *Nat. Photon.* **14**, 623-628 (2020).

29. Yesilkoy, F. *et al.* Ultrasensitive hyperspectral imaging and biodetection enabled by dielectric metasurfaces. *Nat. Photon.* **13**, 390-396 (2019).

30. Sievenpiper, D., Zhang, L. J., Broas, R. F. J., Alexopolous, N. G. & Yablonovitch, E. High-impedance electromagnetic surfaces with a forbidden frequency band. *IEEE Trans. Microw. Theory Techn.* **47**, 2059-2074 (1999).

## Methods

### Experimental samples.

The PhC slabs are fabricated with standard printed circuit board (PCB) technology. Both PhC samples consist of 48 by 48 mushroom-like unit cells (as shown in Fig. 1a). For the slab with the symmetry-protected BIC (the nonsymmetry-protected BICs), the thickness of the dielectric material is 4 mm (5.3 mm). Other geometrical parameters are the same: lattice constant $a = 10$ mm, radius of via $r = 2$ mm, width of the upper copper plate $b = 8$ mm, and thickness of the copper layer 0.018 mm. The relative permittivity of the dielectric material is $3.5 + 0.007i$.

### Numerical simulations.

The band structures, field distributions, coupling constants, and quality factors are numerically calculated with the eigenmode solver of the commercial software COMSOL Multiphysics. In simulations, we set periodic boundary conditions (PBCs) in the $x$ and $y$ directions and perfect match layers (PMLs) in the $z$ direction, and the height of the air box is 10 times the lattice constant. The reflection coefficients are numerically calculated with the frequency domain module of the commercial Computer Simulation Technology (CST) Microwave Studio software. The "unit cell" condition is adopted in the $x$ and $y$ directions. The copper is treated as a perfect electrical conductor (PEC) in simulations.

### Measurements.

All experiments are performed with a customized far-field measurement platform (Linbou FFS). We fix experimental samples on a rotating motor that can be rotated automatically to scan the azimuth angle ($\varphi$). A pair of high-collimated lens antennas (HD-140LHA250+SZJ: 12.4 GHz~18 GHz) are adopted and fixed on two 1.5 m-long arms that rotate around the experimental sample. The polar angle ($\theta$) can be tuned from 6° to 90° with a resolution of 0.5° by rotating two arms. The polarization of the antennas is adjusted by changing the orientation of the short sides (i.e., electric field) of the rectangular waveguide connector. For the $s$-polarized ($p$-polarized) wave, the short sides are perpendicular (parallel) to the ground. All reflection spectra are measured with a vector network analyzer (Ceyear 3672B 10 MHz-26.5 GHz). More details of the experimental setup can be found in the Supplementary Information.

## Acknowledgments

The work at Zhejiang University was sponsored by the National Natural Science Foundation of China (NNSFC) under Grants No. 61625502, No. 62175215, No. 11961141010, No. 61975176, No. 62071418, and No. 61931007, the Top-Notch Young Talents Program of China, the Fundamental Research Funds for the Central Universities, and the Science Challenge Project under Grant No. TZ2018002. The work at Southern University of Science and Technology was sponsored by SUSTech Start-up Grant (Y01236148), Young Thousand Talent Plan of China and the National Natural Science Foundation of China under Grant No. 12104211.

## Author contributions

Y.Y., Z.G., and F. C. conceived the original idea. Y.Y., Z.G., and F.C. designed the structures and the experiments. F. C., and Z.G. conducted the experiments. F.C., Y.Y., Z.G., L.Z., Q.C., Q.Y., R.X., H.C., and L.J. did the theoretical analysis. F.C., Y.Y., W. Y., H.C., and Z.G. wrote the manuscript and interpreted the results. Y.Y., Z.G., H.C., and W.Y. supervised the project. All authors participated in discussions and reviewed the manuscript.

**Competing interests**

The authors declare no competing interests.

**Data availability**

The data that support the plots within this paper and other finding of this study are available from the corresponding author upon reasonable request.

**Code availability**

All relevant code is available from the corresponding author upon reasonable request.

**Additional information**

**Supplementary Information** is available for this paper.

**Reprints and permissions information** is available at http:xxxx.

**Correspondence and requests for materials** should be addressed to Y.Y. or Z.G. or H.C. or W.Y.

**Publisher's note:** xxxx remains neutral with regard to jurisdictional claims in published maps and institutional affiliations.